\documentclass[12pt,a4paper, twocolumn]{article}
\usepackage[hmargin={0.6in,0.6in},vmargin={1in,1in}]{geometry}
\usepackage{pbox}
\usepackage{amsmath,amssymb}
\usepackage{bbm}
\usepackage{changepage}
\usepackage[utf8x]{inputenc} 
\usepackage{textcomp,marvosym} 
\usepackage{cite} 
\usepackage{abstract}
\usepackage[affil-it]{authblk}
\usepackage[right]{lineno} 
\usepackage{microtype}
\usepackage[table]{xcolor} 
\usepackage{array}
\usepackage{marvosym}
\usepackage{tkz-graph}
\usepackage{float}
\usepackage{pgfplots}
\usepackage{pgfplotstable}
\usetikzlibrary{arrows, positioning,automata,shapes}
\usepackage{caption}
\usepackage{subcaption}
\usepackage{natbib}

\title{Evolution and structure of technological systems - An innovation output network 
\footnote{I gratefully acknowledge funding support from the Jan Wallander and Tom Hedelius foundation (grant no W2015-0445:1) and Sweden's governmental agency for innovation systems, Vinnova (grant no 2014-06045). Part of this paper was written during a research visit at INET Oxford. I gratefully acknowledge the hospitality of the Complexity Economics programme. 
}
 }

\author{
Josef Taalbi
}
\affil[1]{\small Department of Economic History, Lund University.}

\date{\today}

\begin{document}
\twocolumn[
 \begin{@twocolumnfalse}

\maketitle

\begin{abstract}
\noindent
This study examines the network of supply and use of significant innovations across industries in Sweden, 1970-2013. It is found that 30\% of innovation patterns can be predicted by network stimulus from backward and forward linkages. The network is hierarchical, characterized by hubs that connect diverse industries in closely knitted communities. To explain the network structure, a preferential weight assignment process is proposed as an adaptation of the classical preferential attachment process to weighted directed networks. The network structure is strongly predicted by this process where historical technological linkages and proximities matter, while human capital flows and economic input-output flows have conflicting effects on link formation. The results are consistent with the idea that innovations emerge in closely connected communities, but suggest that the transformation of technological systems are shaped by technological requirements, imbalances and opportunities that are not straightforwardly related to other proximities. 


\vspace{1cm}
\end{abstract}

 \end{@twocolumnfalse}
]
\saythanks

\section{Introduction}
\label{intro}

A fundamental issue in the study of complex networks is to identify mechanisms that shape the structure and evolution of biological, economic and social networks, such as the Internet, neural networks and scientific citations and collaborations \citep{newman2001, albert2002, dorogovtsev2000, dorogovtsev2002, newman2005}. 
Innovation networks share many often observed features of real networks, such as scale-free degree distributions and non-trivial clustering. Yet, the mechanisms responsible for the structure and evolution of innovation networks are still not well understood. In the field of innovation studies, scholars have for some time theorized the mechanisms that drive the structure and evolution of \emph{technological systems}, i.e. networks of agents involved in the generation and use of technology, which in the presence of radical innovation and momentum can become synergistic clusters \citep{Carlsson1991a}. A large literature suggests the evolution of such technological systems to be driven by innovational complementarities, knowledge spillovers and increased incentives to innovate owing to technological advances upstream \citep{romer1986, David1990, bresnahan1995, Verspagen1997, lipsey2005, acemoglu2016}, or driven by the response of agents to demand and systemic imbalances downstream in the technological system \citep{Hughes1987, Rosenberg1969}.   
Notably, in a recent study, \citet{acemoglu2016} found strong evidence for a role played by upstream network stimulus in the US patent citation network.

So far, such empirical studies have made use of R\&D and patent flows, which does not necessarily capture innovation and, as noted by \citet{acemoglu2016}, might be understood as 'normal science'. Noting that the evolution of technological systems and sociotechnical transitions are driven by the development and diffusion of major or radical innovations \citep{Carlsson1991a, geels2002, geels2005}, there are open questions regarding the role of network stimulus and the mechanisms responsible for the structure and evolution of technological systems. This paper studies these problems through the lens of inter-industrial networks of significant innovations, viz. commercialized inventions that have a non-incremental degree of novelty (compare \citealp{Pavitt1984, taalbi2017}). Through analysis of such an innovation output network for Sweden, 1970-2013, this study examines underlying dynamics that shape innovation patterns over time and the structure of supply and use of innovations across industries. Previous studies on this database \citep{taalbi2017,taalbi2017b} have found that innovation activity is typically the response to problems and opportunities that have emerged elsewhere in the technological system and that these problems and opportunities focus innovation activity in distinct communities. However, no earlier attempt has been made to explain the structure and evolution of the technological system through network analysis.

The present study suggests that network stimulus from industries upstream and downstream explains variations within and between industries and that the evolution of the innovation network is best described as a process where previous ties and proximities in the innovation network largely determine the appearance of new ties. The process is driven by hubs, key suppliers, that connect diverse user industries in stable communities. While there is moderate evidence for a role played by underlying economic, skill or knowledge proximities, the results stress that the bulk of significant innovations are geared towards technological diversification rather than following existing industry interdependencies. Hence, the results are consistent with a view of innovation activity as a process of co-evolution between industries, but emphasize that the formation of technological systems are shaped by technological requirements, imbalances and opportunities that may not be squarely captured through economic or other proximities.  

Section \ref{sec: evol} discusses the theoretical framework for network evolution. Section \ref{sec:methods} discusses the data sources and variables used. Section \ref{sec:results} presents the results, followed by concluding remarks.

\section{Evolution of innovation networks}
\label{sec: evol}
\subsection{Extrinsic and intrinsic mechanisms}

What mechanisms account for the structure and evolution of innovation networks? One may broadly distinguish between mechanisms intrinsic and extrinsic to a network under study.  
Drawing on seminal work \citep{dorogovtsev2000, newman2001, newman2005, albert2002}, the literature on evolving networks offers a few intrinsic candidate mechanisms. 

Foundational insights stem from the notion that different linkage formation mechanisms are reflected in qualitatively different degree distributions. Random graphs generated by the Erd\H{o}s-R{\'e}nyi model with homogenous probability, predicts a binomial degree distribution. Many large real world networks have however been shown to have scale-free distribution in the node degrees $k$ 
 of the form \begin{equation} P(k) \propto k^{-\gamma} \end{equation} where in practice it is often found that $2 \leq \gamma \leq 3$. In the context of growing networks, this has been explained by the preferential attachment of new nodes to nodes that are already well connected \citep{barabasi1999}.  
 The preferential attachment process is well in accordance with the notion in innovation studies that broad technology shifts are characterized by path dependence, increasing returns and the emergence of a persistent core-periphery structure where general-purpose technologies supply most of the innovations across the board \citep{david1985,arthur1987, lipsey2005}. Hence the evolution of innovation networks, like many other networks \citep{barabasi1999, newman2001}, may be a process determined largely by network topology. 

Other mechanisms related to the network topology have however been suggested to be capable of explaining community structure in complex networks, notably based on the notions of triadic closure and node similarity \citep{adamic2003, liben2007, bianconi2014}. In general, the probability that two nodes will become connected may depend on the respective number of neighbors or the indirect paths between nodes. For instance, it is intuitive that two scientists may be more likely to collaborate in the future if they have mutual connections. 
Along these lines, an important perspective is that innovation flows are governed by a process of co-evolution across industries, in the sense that innovations respond to opportunities and problems that appear in closely knitted technological sub-systems \citep{Carlsson1991a, nelson1994, bresnahan1995, taalbi2017}. Co-evolution is stipulated to take place in terms of opportunities provided upstream in the value-chain \citep{bresnahan1995, lipsey2005, acemoglu2016}, but also through problems and imbalances downstream that focus innovation activity \citep{Rosenberg1969, Hughes1987}.  

Innovation studies also advance several reasons why innovation flows could be expected to evolve due to mechanisms \emph{extrinsic} to innovation processes. One hypothesis advanced is that innovation networks in general co-evolve with underlying economic interdependencies and different kinds of proximities, e.g., cognitive, organizational, social, institutional and geographical \citep{boschma2005, boschma2010}. It is for instance plausible that innovation flows follow existing input-output relationships between industries of capital or intermediate goods \citep{debresson1991,debresson1996b}, in line with accounts that stress the role of demand conditions in spurring innovation \citep{schmookler1966,Lundvall1988,VonHippel1988}. Cognitive proximities are also likely to induce innovation, for instance in the sense that innovative problem solution requires a shared knowledge base between the technology being developed and the user application. Similarly, users need complementary absorptive capacity to exploit innovations from other industries \citep{cohen1990, nooteboom2007}. 

The suggested importance of economic and cognitive proximities in innovation processes however comes with caveats, as processes of radical and incremental innovation are fundamentally different phenomena. Firms often innovate in order to branch into fields where they are not economically active, viz. through exploration and diversification \citep{march1991}. As it were, radical innovation may be associated with exploration and ventures into new markets, while more incremental innovations are more likely to follow established paths such as production networks. Moreover, too high (e.g., cognitive) proximities between industries may be negatively associated with innovation. Indeed, previous comparisons of innovation indicators and measures of relatedness have suggested mixed relationships, and that radical innovation is associated with unrelated variety \citep{castaldi2015}.

\subsection{Evolution of a weighted innovation network}
Bringing these perspectives together, the formation of network ties can be viewed as a process in which there are several potential mechanisms at play: preferential attachment, triadic closure and underlying economic, knowledge and technological proximities. The innovation dynamics can be modelled on two levels. It is first important to understand to what extent the pre-existing network structure predict innovation performance. Following \citet{acemoglu2016}, one may model the count of innovations by sector as predicted by network stimulus from backward and forward linkages in previous periods. The details of such a model are described in connection with the empirical results in section \ref{sec:mult}.

Section \ref{sec:drivers} then analyzes the formation process of ties between industries and the structure of the network. To model network evolution, one may start by considering a straightforward framework for link formation where the probability of a new innovation between industry $i$ and industry $j$ depends on the cumulative flow of innovations $w_{ij}$. Intersectoral innovation networks are weighted directed networks, with a fixed number of nodes. Preferential attachment in a setting where direction and weight matter, relates the probability of flows to the in- and out-strength of nodes, defined as $s^{in}_i = \sum_j w_{ji}$ and $s^{out}_{i}=\sum_j w_{ij}$. The process is defined by the notion that the probability that an innovation is supplied an industries $i$ depends on $\frac{s^{out}_i}{\sum w_{ij}}$, and that is used by industry $j$ depends on $\frac{s^{in}_j}{\sum w_{ij}}$. The probability of a new link from industry $i$ to $j$ can then be expressed as the product of the two terms $\Pi_w =\alpha \frac{s^{out}_{i}s^{in}_{j}}{\left(\sum w_{ij}\right)^2} $.

It is however advantageous to introduce node heterogeneity to the preferential attachment process. In this conceptualization, the probability that innovations are supplied by industries $i$ depends on $\frac{s^{out}_i}{\sum w_{ij}}$, but, given the industry of supply, innovations are more likely to be used by industries which are typical users of industry $i$, hence depending on $\frac{w_{ij}}{s^{out}_i}$. The product of these two terms gives the probability of a link between $i$ and $j$ as 

\begin{equation} 
\label{eq:main}
\Pi_w \propto \frac{w_{ij}}{\sum w_{ij}}
\end{equation}

Clearly, introducing node heterogeneity results in a process of preferential assignment of innovations to industry pairs (edges), rather than to particular nodes. In the remainder of this study, "preferential attachment" is used to refer to this type of process unless stated otherwise. When conceptual separation from the classical model is needed the perhaps more accurate label "preferential weight assignment" (PWA) will be used.

Since at the outset no edges have innovations, the process is appropriately modelled as a mixed process of preferential attachment and random assignment of innovations to unused edges. A sensible generalization is to let the preferential attachment process be potentially non-linear through a parameter $\lambda$, leading us to specify the probability for an edge with weight $w$ as   

\begin{equation}
\label{eq:constant}
\Pi_w =(1-\alpha)\delta_{w,0} +\alpha \frac{w^\lambda}{\sum w^\lambda N_w}
\end{equation} where $\delta_{w,0}$ is $1$ for $w=0$, otherwise $0$ and $N_w$ is the number of edges with weight $w$. For $\alpha = 1$, the model reduces to a pure preferential attachment process. 

The linear model, i.e. when $\lambda=1$ is our baseline model.   
The distribution resulting from equation \ref{eq:constant} however depends on the parameter $\lambda$ and whether the number of edges of the network are finite or growing \citep{bagrow2008}. 
In our case (section \ref{sec:methods}), the network size is fixed but the time lapsed is smaller than the number of edges. In this scenario, the network can be treated as growing, with $\alpha$ as a historically observed parameter. 
Appendix \ref{sec:distribution} derives and details the relevant weight distributions for linear and sub-linear attachment kernels. For the baseline linear case $\lambda =1$, weights follow a Pareto distribution $P(w) \propto w^{-1-1/\alpha}$.  For the sub-linear case $\lambda<1$, weights follow a stretched exponential.\footnote{Notice that our model only contains a random process for edges with weight zero. Compare the model set up by \citet{bagrow2008} for fixed size networks, which includes a random homogeneous process for \emph{all} nodes. They show that with a fixed number of nodes $A$, the limiting distribution of the process 
 depends on the parameter $\alpha$ and the time lapsed.  
In the asymptotic limit, the system has two attractors. If $\alpha<\frac{1}{2}$, the distribution approaches a Gaussian of width $\frac{t}{[1-2\alpha]A}$. If $\alpha>\frac{1}{2}$, the distribution retains the power-law distribution in the tail. }

Equation \ref{eq:constant} describes our baseline model, used to make predictions on weight distribution. However, it is also desirable to test for the effect of exogenous networks $z_{ij}$ capturing (possibly evolving) proximities between industries however measured. This gives a test equation

\begin{equation} 
\Pi_w  = \beta_0 +\beta_1 \frac{w_{ij}}{\sum w_{ij}}+\sum_l \beta_l z_{ij,l}
\end{equation} 

It is also desirable to test for other link formation mechanisms, such as triadic closure and the effect of indirect links, using node similarity metrics $S_{ij}$. These are tested using link prediction methods in section \ref{sec:link}.

\section{Materials and methods}
\label{sec:methods}

Innovation networks, in the sense of innovation flows between sectors, have been studied in a large number of empirical works, using a variety of measurements of innovation: patent data (\citealp{Scherer1982, Verspagen1997, vanMeijl1997, nomaler2008, nomaler2012, acemoglu2016}), R \& D flows \citep{leoncini1996, Leoncini2003, Montresor2008, hauknes2009} and innovation output data \citep{debresson1978, Robson1988, debresson1996}. 
The innovation network used in this study stems from a Swedish innovation output database \citep{sjoo2013, taalbi2017}, assembled through a literature-based innovation output methodology \citep{Kleinknecht1993b}. 15 trade journals, covering the manufacturing and ICT service sectors, were screened for significantly improved products, commercialized on a market. The innovations collected were subject to a criterion of commercial status and novelty according to independently edited journal articles. Hence, innovations are not incremental or mere inventions without economic or societal interest, but significant and novel to the industry. 

The innovation network is constructed by mapping the number of innovations flowing from product groups to the sectors in which the innovation is used or explicitly intended to be used. This refers strictly to the commercial use of the innovation, as distinguished from patent citation networks that map knowledge flows between technologies. The product innovations found in the journal articles were categorized according to the Swedish Industrial Classification system 2002 (SNI 2002) corresponding to ISIC Rev. 3. For each innovation, user industries were likewise assigned according to ISIC Rev. 3, at the lowest level of detail possible.  

In practice, the innovation output network used in this study is a weighted directed network with industries as nodes and innovation counts as edges. The full detail innovation network is a $98 \times 98$ matrix, with 9,604 possible entries per year. Since there is a selection on the source industries of innovations, being in practice limited to manufacturing and ICT services, regressions include flows from these sectors to others, in practice 6370 network edges. Since any innovation may have several user industries, a weighting procedure was constructed so that each innovation is only counted once. Hence an innovation that has $n$ user industries will have $n$ linkages of weight $1/n$. Innovations may be partially for final consumption or have a general-purpose character. These user categories have not been included in the network analysis. Out of 3685 innovations, the sum total of between industry flows was $3248.2$. 

From this data, the yearly flows of innovations $\Delta w_{ij}$ and the cumulative flow of innovations between industries $w_{ij}$ are constructed. To account for node proximities, this data is also used to construct a battery of similarity metrics. A first set of measures aim to capture triadic closure: if two nodes $i$ and $j$ have links to a third node, they are likely to obtain links too. These local metrics are based on the number of common neighbors of two nodes $i$ and $j$

 \begin{equation}
\sum_{k \in \Gamma(i)\cap \Gamma(j)} w_{ik}+w_{jk} 
\end{equation} with $\Gamma(i)$ being the set of neighboring nodes of $i$, and $k$ an index of neighboring nodes.

The Jaccard similarity metric normalizes the effect of neighborhood size 

\begin{equation}
\sum_{k \in \Gamma(i)\cap \Gamma(j)} \frac{ w_{ik}+w_{jk} }{\sum_{l \in \Gamma(i)} w_{il}+\sum_{m \in \Gamma(i) }w_{im}} 
\end{equation}

Since rare connections may be more indicative of where new links emerge, the Adamic-Adar metric (\citeyear{adamic2003}) normalizes the metric by letting a common neighbor $k$ be weighted by the rarity of relationships between other nodes and $k$
\begin{equation}
\sum_{k \in \Gamma(i)\cap \Gamma(j)} \frac{ w_{ik}+w_{jk} }{\log \left( 1+ \sum_{m \in \Gamma(k)} w_{km} \right) } 
\end{equation}

Global metrics are based on the overall network. In this study we use the Katz metric (\citeyear{katz1953}), which is the sum of all walks that exist between nodes, weighted by a parameter $\beta^l$ that decreases with path length $l$. In compact matrix notation the metric is defined as

\begin{equation}
\mathbf{S}^{\text{katz}}=\sum_{l=1}^{\infty} \beta^l \mathbf{W}^l = {(\mathbf{I}-\beta \mathbf{W})^{-1}}-\mathbf{I}
\end{equation}
with $\mathbf{I}$ the identity matrix, $\mathbf{S}^{\text{katz}}=||S_{ij}^{\text{katz}}||$, $\mathbf{W}=||\dfrac{w_{ij}}{\sum w_{ij}}||$ and $1/\beta$ being greater than the largest eigenvalue of $\mathbf{W}$. This metric has the advantage that it can be decomposed into the matrix of direct links $\mathbf{W}$ and indirect (second and higher order) links $\mathbf{S}^{\text{katz}}-\beta \mathbf{W}$. 

To measure underlying economic proximities, the Leontief inverse is used, constructed from Input-Output tables 1995-2011 (OECD). This is calculated from the accounting identity (in matrix notation) 
$\mathbf{Ax}+\mathbf{y}=\mathbf{x}$, with $\mathbf{A}$ an $N$ by $N$ input coefficient matrix, and $\mathbf{x}$, $\mathbf{y}$ are $1 \times N$ output and value added vectors respectively. The Leontief inverse matrix is defined as the marginal effect of a change in value added of an industry on the demand for goods, in compact matrix form:
\begin{equation}
\mathbf{L}= (\mathbf{I}-\mathbf{A})^{-1}
\end{equation}

We also follow earlier studies \citep{leoncini1996, montresor2009} and use the R\&D expenditure embodied in direct and indirect economic good requirements between sectors as a measure of knowledge flows, viz. knowledge proximities between industries. This measure assumes that innovation is embodied in the exchange of intermediate goods, calculating a network $\mathbf{R}$ from the intra-sectoral R\&D  expenditures and Leontief inverse 

\begin{equation}
\mathbf{R} = \mathbf{\hat{r}}\mathbf{\hat{q}}^{-1} \left(\mathbf{I}-\mathbf{A} \right)^{-1} \mathbf{\hat{y}}
\end{equation} where $\mathbf{\hat{r}}$, $\mathbf{\hat{q}}$ and $\mathbf{\hat{y}}$ are diagonalized vectors of R\&D expenditure, output and final demand respectively. 

As an additional measure of economic proximities this study also employs the "Los index" \citep{los2000}, which departs from the idea that the relatedness between industries is expressed through the similarity of the input goods used. The technological relatedness is then measured as the cosine similarity between a pair of input coefficient vectors $a_{ik}$ and $a_{jk}$ according to

\begin{equation}
\omega_{ij} = \frac{\sum_{k=1}^n a_{ik} \cdot a_{jk}}{\sqrt{\sum_{k=1}^n a^2_{ik} \cdot \sum_{k=1}^n a^2_{jk}}}
\end{equation} with $i,j,k \in \left\lbrace 1,..., n \right\rbrace$ as sector indices.

To measure knowledge-base proximities, a skill-relatedness measure is used, constructed by \citet{neffke2011}. This measure is based on human capital flows between industries at the 4-digit level for Sweden, constructed for the years 2004-2007. In addition, the model estimations also use value added 1970-2013 as a control variable for "gravity" between industry $i$ and $j$, 

\begin{equation}
F_{ij}=g \frac{m_i m_j}{m^2}
\end{equation}
where $m_i$, $m_j$ is the value added of industry $i$ and $j$ respectively, $m$ is total value added and $g$ is a parameter.

\section{Results} \label{sec:results}
\subsection{Network structure}
\label{sec:structure}
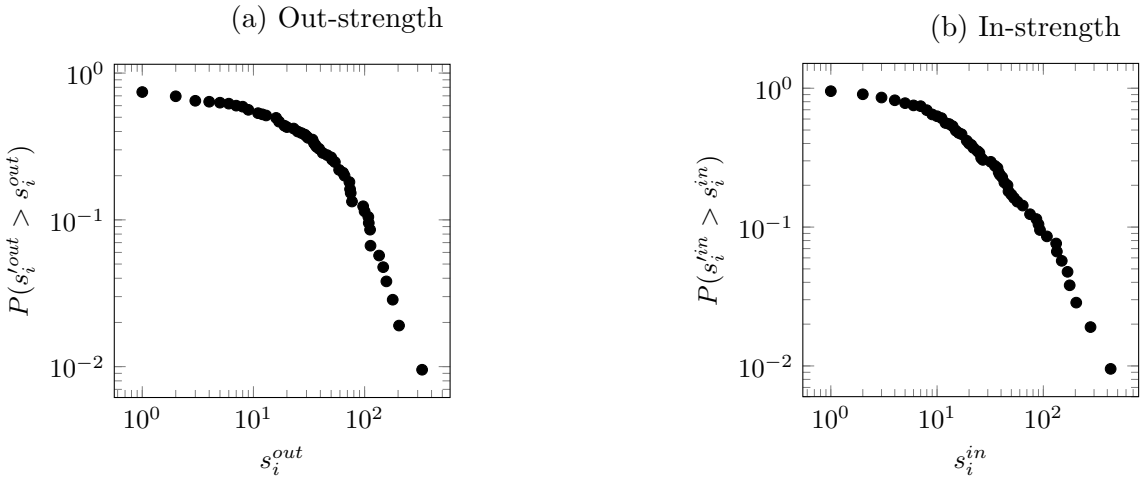
\begin{figure*}
\caption{Out- and in-strength distribution, 1970-2013}
\label{fig:outindegree}
\footnotesize
\begin{subfigure}[b]{0.50\textwidth}
\caption{Out-strength}
\pgfplotstableread{Q.txt} \Q
\begin{tikzpicture} 
\begin{axis}  [height = 6cm, width = 6cm, xmin = 0, xmode = log, ymode = log, xticklabel style={/pgf/number format/fixed}, ylabel = $P(s'^{out}_i>s^{out}_i)$, xlabel = $s^{out}_i$]
\addplot [black, only marks] table[y = Freq, x=deg] from \Q;
\end{axis}
\end{tikzpicture}
\end{subfigure}
\begin{subfigure}[b]{0.50\textwidth}
\caption{In-strength}
\label{logdegree}
\footnotesize
\pgfplotstableread{Qin.txt} \Q
\begin{tikzpicture} 
\begin{axis}  [height = 6cm, width = 6cm, xmode = log, ymode = log, xticklabel style={/pgf/number format/fixed}, ylabel = $P(s'^{in}_i>s^{in}_i)$, xlabel = $s^{in}_i$]
\addplot [black, only marks] table[y = Freq, x=deg] from \Q;
\end{axis}
\end{tikzpicture}
\end{subfigure}
\end{figure*}

\begin{figure}[H]
\caption{Cumulative distribution of innovation flows $w_{ij}$ (log-log), 1970-2013 and theoretically predicted distributions obtained by using values of the fitted parameters (see Tables \ref{tab:results}, model 1, \ref{tab:logs1} and Appendix \ref{sec:distribution})}
\label{fig:weights}
\footnotesize
\pgfplotstableread{Q2.txt} \Q
\begin{tikzpicture} 
\begin{axis}  [height = 7cm, width = 8cm, xmode = log, ymode = log, xticklabel style={/pgf/number format/fixed},
ylabel = $P(w'_{ij}>w)$, xlabel = $w$]
\addplot [black, only marks] table[y = Freq, x =deg] from \Q;
\addlegendentry{$P(w'_{ij}>w)$}        
 \addplot[red, domain=1:100] {(0.25*(x)^(-1/0.625))}; 
 \addlegendentry{Linear}   
\addplot[blue, domain=1:85] {0.3*exp(-0.5963088/(1-0.6492)*x^(1-0.6492)};  
\addlegendentry{Sub-linear}   
\end{axis}
\end{tikzpicture}
\end{figure}
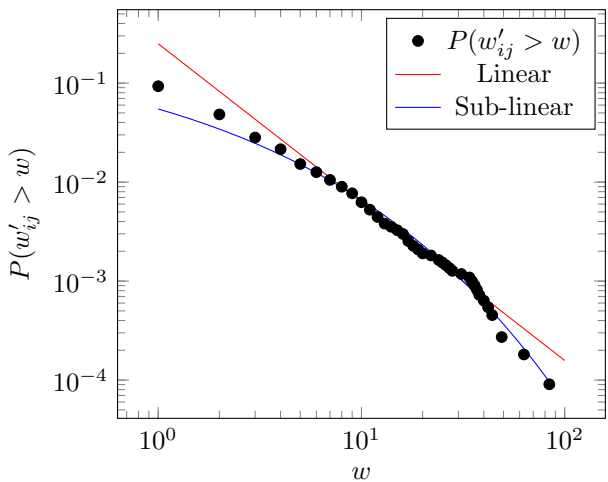

The network of supply and use of innovation objects reveals several structural properties of interest:  stability, hierarchy and asymmetry. Overall the network is very stable over time: Pearson correlations between 5-year periods range between 0.46 and 0.62 (see Table \ref{tab:correlation}). Figures \ref{fig:heatmap1}-\ref{fig:heatmap3} in Appendix \ref{sec:network} highlight the heterogeneity and asymmetric character of innovation flows between 29 aggregate sectors. Some industries, mainly machinery, ICTs and software, tend to be major suppliers, supplying to a broad range of industries, while others (e.g., construction and transport services) are almost exclusively users of innovation. Figure \ref{fig:outindegree} shows an exponential weight distribution in the total out- and in-strength of industries. The weight distribution (Figure \ref{fig:weights}), lies closer to a power-law predicted by a linear preferential attachment process, but is better fitted as a stretched exponential, as predicted by a sub-linear rate of attachment (see Table \ref{tab:results} and Appendix \ref{sec:distribution}). 

\begin{figure}
\caption{Community structure of the network of innovations, 1970-2013.}
\label{fig:network}
\centering
\includegraphics[width=8cm, height = 8cm, 
]{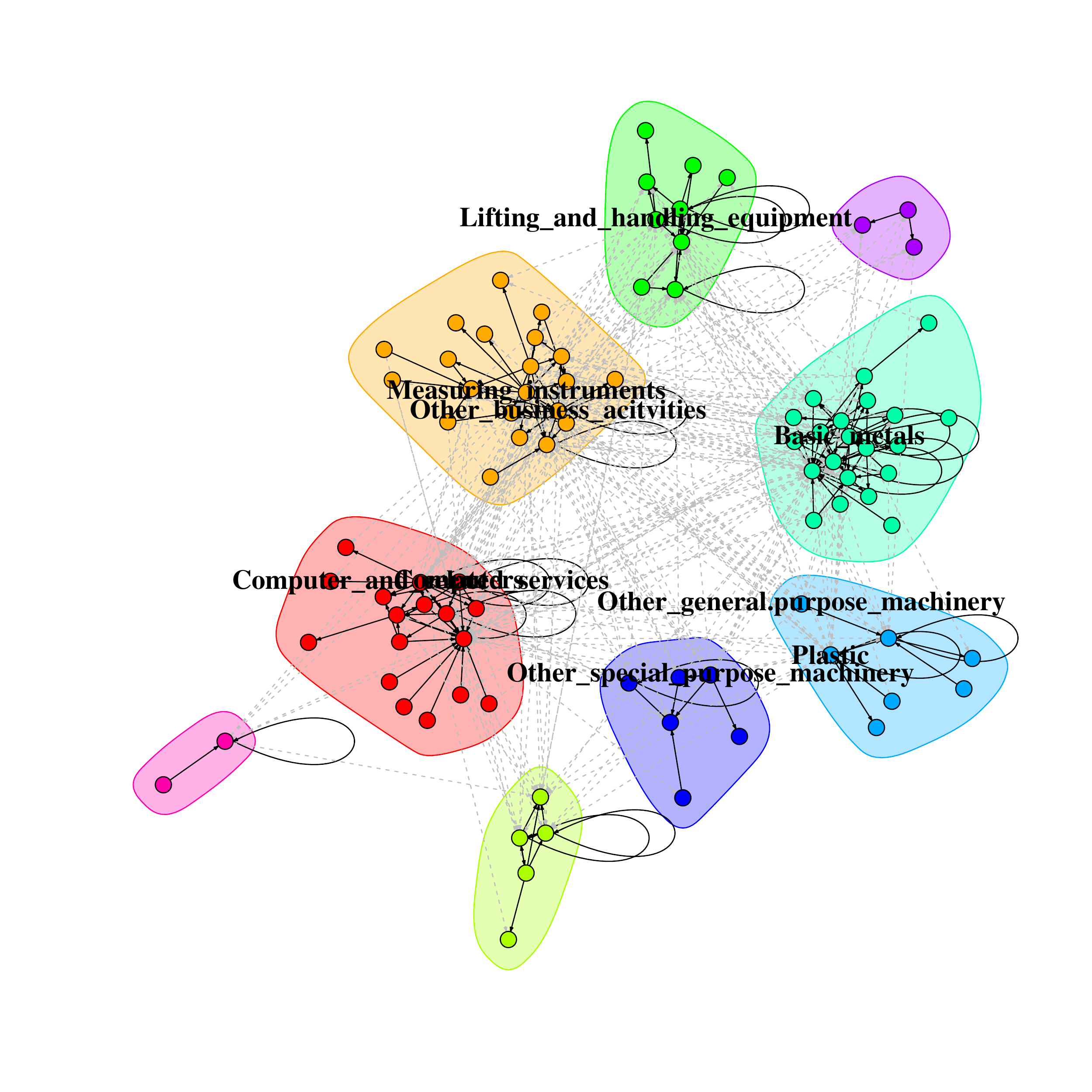}
\tiny
\subcaption*{Layout by communities, applying the fast greedy community detection algorithm \citep{clauset2004}. For further discussion, see \citet{taalbi2017}.}
\end{figure}

\begin{figure}[H]
\centering
\caption{Local clustering coefficients. ICT industries in blue. Dashed line $k^{-1}$}
\label{fig:clustering}
\footnotesize
\pgfplotstableread{clusteringcolor.txt} \C
\begin{tikzpicture} 
\begin{axis}  [height = 7cm, width = 8cm, xmode = log, ymode = log, xticklabel style={/pgf/number format/fixed}, ymin = 0.2, ymax=1.2,
xmax=200,
 ylabel = Clustering coefficient $C(k)$, xlabel = Node degree $k$]
]
\addplot [black, only marks, mark = o] table[y = C, x=deg] from \C;
\addplot [blue, only marks, mark = *] table[y = ICTC, x=ICTdeg] from \C;
\addplot[domain=1:200,dashed,thick]{25*x^-1};
\end{axis}
\end{tikzpicture}
\end{figure}
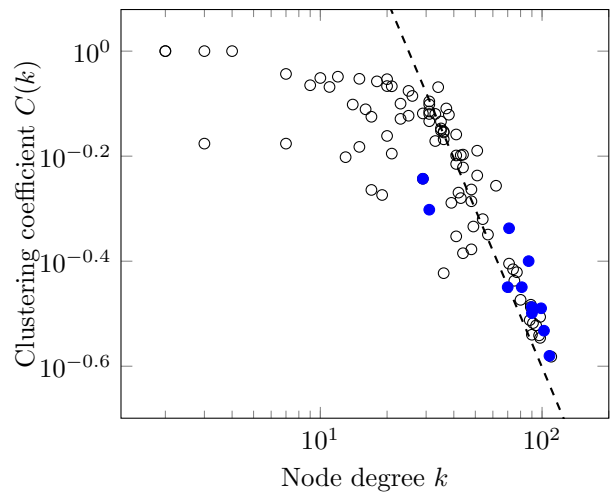

Secondly, it is possible to observe that the network is hierarchical and has a significant community structure within the network (see also \citealp{taalbi2017}). The network has a modular structure similar to the hierarchical model suggested by \citet{ravasz2003}. The overall probability that two neighbors of a node are themselves neighbors is a sizeable 0.608. 
Moreover, the local clustering coefficient $C$, measuring the average probability that two neighbors of a node are neighbors, scales like $C \sim k^{-1}$ in the tail (Figure \ref{fig:clustering}). This indicates a hierarchy in which low degree nodes belong generally to well interconnected communities (high clustering coefficient), while hubs connect many nodes that are not directly connected (low clustering coefficient) \citep{ravasz2003}. In our case, major innovation suppliers such as ICTs act as hubs that connect user industries in a more or less star-like structure (compare \citealp{taalbi2017origins}). More specifically, the interdependencies in the Swedish network of innovations can be described in terms of ten subgroups (Figure \ref{fig:network}) centered on pulp and paper, foodstuff, ICTs, automotive vehicles and land transportation, medical equipment and healthcare, forestry, construction, military defense and military suppliers, electricity and textiles (see \citealp{taalbi2017} and \citealp{taalbi2017origins} for further discussion). 

\subsection{Network stimulus effects}
\label{sec:mult}
Let us first turn to what role the network structure has for innovation performance over time. To construct predicted innovations in a five-year period from forward and backward flows, we combine the pre-existing network with the number of innovations that were launched in the previous five-year period. Let $X_{i,T}$ be the number of innovations in industry $i$ in period $T$. Expected innovation counts from forward linkages are calculated as   

\begin{equation}
\widehat{X}^{f}_{i,T} = \sum_j \dfrac{w_{ji}}{s^{out}_j} X_{j,T-1}
\end{equation} and expected innovations from backward linkages as
\begin{equation}
\widehat{X}^{b}_{i,T} = \sum_j \dfrac{w_{ij}}{s^{in}_j} X_{j,T-1}
\end{equation} where $w_{ij}$ is, as before, the cumulative flow of innovations from industry $i$ to industry $j$ and $s^{in}_j$ and $s^{out}_j$ are in- and out-strength of node $j$. Expected innovations from forward flows are thus specified by combining the innovations in the previous five-year period with pre-existing forward flows. Expected innovations from backward flows are similarly specified by combining innovation in the previous five-year period with the pre-existing backward flows. 

These formulations assume that the extent to which innovations respond to opportunities upstream or problems and demands downstream is related to the number of innovations in the previous five-year period. This is motivated from the point of view that industries with many innovations will engender new opportunities \citep{bresnahan1995, lipsey2005, acemoglu2016}, while also drawing attention to technological imbalances and critical problems that appear in the process of technological change (cf. \citealp{Rosenberg1969, Hughes1987, taalbi2017b}).  

Our results (Table \ref{tab:Techmult}-\ref{tab:Techmult2}) are in line with both notions. The analysis focuses on 44 industries that have launched at least one innovation biannually and 29 that launched at least one innovation per year. In a pooled regression (models 1-2), $30\%$ of the variation in innovation is explained by forward linkages and backward linkages in the previous five-year period. The coefficients indicate that an increase in the predicted innovations of $1\%$ increases the number of innovations with $0.38\%$ and $0.34\%$. Since these results may to some degree reflect common innovation trends or persistence of innovation counts across industries, a fixed-effects regression is also specified. In a fixed effects model, backward linkages on their own explain $6.4\%$ of variations within industries, while including time dummies and forward linkages account for $20.7\%$ of within-industry variations. Using the 29 most innovative industries our models explain $25\%$ of variations in innovation launches overall. Backward linkages account for $12.8\%$ in a fixed effects formulation. 

These results are only somewhat less striking than the results from USPTO patent citation networks, reporting network stimulus to account for $55\%$ in a pooled model and $14\%$ in a fixed-effects model \citep{acemoglu2016}. Importantly however, in our case, there is also evidence for a role played by backward linkages in predicting the locus of innovation activity, and in particular in predicting within-industry variations. 

\begin{table*}
\caption{Regression analysis of network stimulus effects.}
\begin{subtable}{.5\textwidth}
\caption{44 industries with at least 1 innovation biannually.}
\label{tab:Techmult}
\centering 
\footnotesize
\begin{tabular}{lcccc} \hline
 & (1) & (2) & (3) & (4) \\
VARIABLES & Pooled & Clustered standard errors & Fixed effects & Year and fixed effects \\ \hline
 &  &  &  &  \\
Forward linkages & 0.383*** & 0.383*** &  & 0.216*** \\
 & (0.0384) & (0.0493) &  & (0.0627) \\
Backward linkages & 0.344*** & 0.344*** & 0.164*** & 0.146*** \\
 & (0.0316) & (0.0487) & (0.0550) & (0.0501) \\
Time dummies &No & No &No & Yes \\
Constant & 0.809*** & 0.809*** & 1.469*** & 1.841*** \\
 & (0.105) & (0.140) & (0.130) & (0.174) \\
 &  &  &  &  \\
Observations & 319 & 319 & 321 & 319 \\
R-squared & 0.300 & 0.300 & 0.064 & 0.207 \\
\hline
\multicolumn{5}{c}{ Robust standard errors in parentheses} \\
\multicolumn{5}{c}{ *** p$<$0.01, ** p$<$0.05, * p$<$0.1} \\
\end{tabular}
\end{subtable}

\begin{subtable}{.5\textwidth}
\caption{29 industries with at least 1 innovation per year.}
\label{tab:Techmult2}
\centering 
\footnotesize
\begin{tabular}{lcccc} \hline
 & (1) & (2) & (3) & (4) \\
VARIABLES & Pooled & Clustered standard errors & Fixed effects & Year and fixed effects\\ \hline
 &  &  &  &  \\
Forward linkages & 0.211*** & 0.211*** &  & 0.187*** \\
 & (0.0473) & (0.0563) &  & (0.0594) \\
Backward linkages & 0.288*** & 0.288*** & 0.240*** & 0.161** \\
 & (0.0392) & (0.0575) & (0.0699) & (0.0664) \\
Time dummies &No & No &No & Yes \\
Constant & 1.296*** & 1.296*** & 1.600*** & 2.054*** \\
 & (0.144) & (0.182) & (0.172) & (0.262) \\
 &  &  &  &  \\
Observations & 218 & 218 & 220 & 218 \\
R-squared & 0.251 & 0.251 & 0.128 & 0.279 \\
\hline
\multicolumn{5}{c}{ Robust standard errors in parentheses} \\
\multicolumn{5}{c}{ *** p$<$0.01, ** p$<$0.05, * p$<$0.1} \\
\end{tabular}
\end{subtable}
\end{table*}

\subsection{Drivers of network evolution}
\label{sec:drivers}
Next we turn to explain the inter-industry tie formation process. The previous results suggest something in between a stretched exponential and a power-law distribution of the innovation flows $w_{ij}$ and out- and in-strength (Figures \ref{fig:outindegree}-\ref{fig:weights}). These results are consistent with a preferential attachment mechanism. 

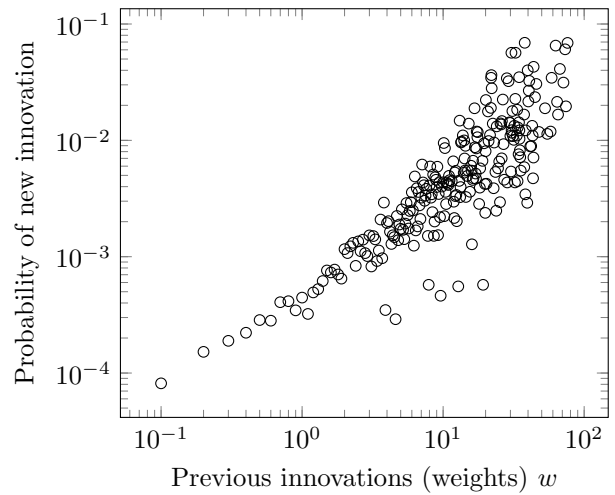
\begin{figure}[H]
\centering
\caption{The probability of an innovation between industries by previous innovations $w_{ij}=w$, 1970-2013.}
\label{fig:relplot}
\footnotesize
\pgfplotstableread{relprob.txt} \C
\begin{tikzpicture} 
\begin{axis}  [height = 7cm, width = 8cm, xmode = log, 
ymode = log, 
xticklabel style={/pgf/number format/fixed}, 
 ylabel = Probability of new innovation, xlabel = Previous innovations (weights) $w$]
]
\addplot [black, only marks, mark = o] table[y = x, x=w] from \C;
\end{axis}
\end{tikzpicture}
\end{figure}

To give a first intuition, Figure \ref{fig:relplot} plots the estimated probability of receiving a new link against the number of previous innovations (cf \citealp{newman2001}), calculated as the average share of innovations for industry pairs with a number $w$ of previous innovations for the period.  
There is clearly a dependence of the linkage formation process on previous ties.

In order to construct empirical models, we note that the (expected) number of new innovations can be linked to equation \ref{eq:constant} through

\begin{equation}
\label{eq:main2}
\Delta w_{ij} = \Delta w \Pi_w
\end{equation}
where innovations arrive at a rate $\Delta w$ per calendar year. To carry out formal tests, equations \ref{eq:constant} and \ref{eq:main2} are combined, leading to the linear baseline model

\begin{equation}
\frac{\Delta w_{ij}}{\Delta w} = \hat{\beta_0}+\hat{\beta_1} \frac{w_{ij}}{\sum_{ij} w_{ij}} + \sum_l \hat{\beta_l} z_{ij,l} +\epsilon_{ij}  
\end{equation} where $z_{ij,l}$ are measures of underlying proximities, and Greek lower case letters are estimated parameters. In inter-industrial innovation networks, one would suspect that the formation of strong ties is in part driven by industrial structure and underlying proximities, but also by previous ties.

\begin{table*}
\caption{Regression results. Dependent variable: relative innovation flows.}
\label{tab:results}
\footnotesize
\centering
\begin{tabular}{lccccc} \hline
 & (1) & (2) & (3) & (4) & (5) \\
 &  & Proximities & Cumulative & Base model & 1995-2011 \\
VARIABLES & Base model & 1995-2011 & 1995-2011 & fixed effects & fixed effects \\ \hline
 &  &  &  &  &  \\
Pref. Att. & 0.625*** &  & 0.693*** & 0.604*** & 0.682*** \\
 & (0.0276) &  & (0.0573) & (0.0296) & (0.0616) \\
Los index &  & -0.000178*** & -2.54e-05 &  & -4.07e-05 \\
 &  & (2.92e-05) & (2.45e-05) &  & (2.78e-05) \\
Skill-relatedness &  & 2.54e-06*** & 3.16e-07 &  & 7.57e-07 \\
 &  & (9.62e-07) & (9.21e-07) &  & (9.20e-07) \\
Gravity &  & 0.172*** & 0.0277 &  & -0.0439 \\
 &  & (0.0284) & (0.0237) &  & (0.0369) \\
Leontief &  & 9.40e-05 & -5.44e-05 &  & 3.62e-05 \\
 &  & (0.000109) & (0.000109) &  & (0.000146) \\
R\&D flows &  & 3.59e-07** & 3.04e-07** &  & 4.53e-07*** \\
 &  & (1.51e-07) & (1.28e-07) &  & (1.34e-07) \\
Constant & 4.02e-05*** & 0.000173*** & 2.09e-05 & 4.05e-05 & -7.41e-05* \\
 & (3.50e-06) & (1.49e-05) & (1.41e-05) & (3.32e-05) & (4.21e-05) \\
 &  &  &  &  &  \\
Observations & 273,910 & 71,370 & 71,370 & 273,910 & 71,370 \\
R-squared & 0.150 & 0.003 & 0.155 & 0.155 & 0.168 \\

\hline
\multicolumn{6}{c}{ Robust standard errors in parentheses} \\
\multicolumn{6}{c}{ *** p$<$0.01, ** p$<$0.05, * p$<$0.1} \\
\end{tabular}
\end{table*}

\begin{table*}
\caption{Logit regression (log odds). 
 Dependent variable: edge (Y/N)}
\label{tab:logit}
\footnotesize
\centering
\begin{tabular}{lccccc} \hline
 & (1) & (2) & (3) & (4) & (5) \\
 &  & Proximities & Cumulative & Base model & 1995-2011 \\
VARIABLES & Base model & 1995-2011 & 1995-2011 & fixed effects & fixed effects \\ \hline
 &  &  &  &  &  \\
Pref. Att. & 0.274*** &  & 0.352*** & 0.490*** & 0.617*** \\
 & (0.00875) &  & (0.0148) & (0.0140) & (0.0230) \\
Los index &  & 0.341*** & 0.586*** &  & -0.0303 \\
 &  & (0.0948) & (0.102) &  & (0.152) \\
Skill-relatedness &  & 0.0149*** & 0.0121*** &  & 0.00822** \\
 &  & (0.00160) & (0.00266) &  & (0.00355) \\
Gravity &  & 502.7*** & 215.2*** &  & -36.16 \\
 &  & (33.87) & (39.99) &  & (110.3) \\
R\&D flows  &  & 0.00124*** & 0.00134*** &  & 0.000443** \\
 &  & (0.000138) & (0.000164) &  & (0.000180) \\
Leontief &  & 1.410*** & 1.646*** &  & -1.503*** \\
 &  & (0.188) & (0.540) &  & (0.381) \\
Constant & 0.102 & -3.777*** & 0.394*** & -0.483 & 0.695 \\
 & (0.0707) & (0.0507) & (0.128) & (0.389) & (0.514) \\
 &  &  &  &  &  \\
Observations & 59,024 & 71,370 & 21,317 & 58,660 & 21,149 \\
 R-squared & 0.0318 & 0.0238 & 0.0620 & 0.139 & 0.166 \\ \hline
\multicolumn{6}{c}{ Robust standard errors in parentheses} \\
\multicolumn{6}{c}{ *** p$<$0.01, ** p$<$0.05, * p$<$0.1} \\
\end{tabular}
\end{table*}

\begin{table*}
\caption{Regression results, non-linear models.}
\label{tab:logs}
\footnotesize
\begin{subtable}{.6\textwidth}
\centering
\caption{Two-step estimation of the parameters of a general preferential attachment process (equation \ref{eq:constant})}
\label{tab:logs1}
\footnotesize 
\begin{tabular}{lcc} \hline
VARIABLES & Step 1 & Step 2 \\ \hline
 &  &  \\
$\hat{\lambda}$ & 0.649*** &  \\
 & (0.00827) &  \\
$\hat{\alpha}$ & & 0.855*** \\
 &  & (0.0044) \\
 &  &  \\
Constant & -1.707*** &  4.54e-06***\\
 & (0.0171) & (2.63e-06) \\
 &  &  \\
Observations & 5,861 & 273,910 \\
 R-squared & 0.512 & 0.121 \\ 
 \hline
\multicolumn{3}{c}{ Standard errors in parentheses} \\
\multicolumn{3}{c}{ *** p$<$0.01, ** p$<$0.05, * p$<$0.1} \\
\end{tabular}
\end{subtable}
\vfill
\begin{subtable}{.6\textwidth}
\caption{Log-log estimations. Dependent variable: relative innovation flows (logs)}
\label{tab:logs2}
\begin{tabular}{lccccc} \hline
 & (1) & (2) & (3) & (4) & (5) \\
 &  & Proximities & Cumulative & Base model & 1995-2011 \\
VARIABLES & Base model & 1995-2011 & 1995-2011 & fixed effects & fixed effects \\ \hline
 &  &  &  &  &  \\
Pref. Att. & 0.675*** &  & 0.641*** & 0.120*** & 0.0878*** \\
 & (0.00712) &  & (0.0166) & (0.0136) & (0.0243) \\
Los index &  & -2.634*** & -1.162*** &  & 0.0581 \\
 &  & (0.153) & (0.132) &  & (0.132) \\
Skill-relatedness &  & -0.00712** & -0.00424* &  & -0.000609 \\
 &  & (0.00299) & (0.00235) &  & (0.00255) \\
Gravity &  & 700.4*** & -0.517 &  & -146.1 \\
 &  & (71.77) & (75.74) &  & (109.7) \\
Leontief &  & -1.403 & -1.827** &  & 0.700 \\
 &  & (0.971) & (0.806) &  & (0.559) \\
R\&D flows  &  & -0.00111*** & -0.00108*** &  & -0.000103 \\
 &  & (0.000378) & (0.000263) &  & (0.000194) \\
Constant & -1.156*** & -5.006*** & -0.512*** & -3.564*** & -4.047*** \\
 & (0.0556) & (0.104) & (0.117) & (0.242) & (0.585) \\
 &  &  &  &  &  \\
Observations & 5,861 & 2,032 & 1,708 & 5,861 & 1,708 \\
 R-squared & 0.565 & 0.153 & 0.584 & 0.818 & 0.849 \\ \hline
 \multicolumn{6}{c}{ Robust standard errors in parentheses} \\
\multicolumn{6}{c}{ *** p$<$0.01, ** p$<$0.05, * p$<$0.1} \\
\end{tabular}
\end{subtable}
\end{table*}

Table \ref{tab:results} reports results from a linear model. The models report robust standard errors. The first estimation is a pooled regression, showing a statistically significant coefficient $\alpha$ for $w_{ij}$ of 0.645. The theoretically predicted cumulative distribution is $P(w'_{ij}>w) \propto  w^{-1/\alpha}$, in our case, $w^{-1.6}$, shown in Figure \ref{fig:weights}. In this model, cumulative innovations explain $15\%$ of the variation in innovation flows. Columns 2-3 report results from including skill-relatedness, input-output proximities and the "gravity" of production volumes. Skill-relatedness and production volumes have significant coefficients but little explanatory power. Columns 4-5 include time dummies and fixed effects for nodes $i$ (sender) and $j$ (receiver). 

This simple approach can be further analyzed in two separate regressions: one for the mere presence of ties and a regression that allows a non-linear dependence of innovation flows on previous ties.  
Table \ref{tab:logit} reports results from a logistic regression where the dependent variable is a binary variable whether there is an innovation between $i$ and $j$. Underlying proximities here have positive and significant effects on the likelihood of an edge between two industries, which is in line with expectations. These proximities seem to mostly capture between-effects. When including fixed effects for nodes $i$ (sender) and $j$ (receiver) only skill-relatedness and embodied knowledge flows have a positive impact.

The most general specification allows for a non-linear attachment rate depending on the parameter $\lambda$. Our model estimated in a two-step regression. First $\lambda$ is estimated using $\log \Delta w_{ij}= \hat{\beta}_0+ \hat{ \lambda } \log w_{ij}$.  Then relative innovation flows $\dfrac{\Delta w_{ij}}{\Delta w}$ are regressed on $\frac{w_{ij}^{\hat{ \lambda }}}{\sum w_{ij}^{\hat{ \lambda } } } $ 
The coefficients are shown in Table \ref{tab:logs1} and the theoretically predicted weight distribution is shown in Figure \ref{fig:weights}. There is clearly a better fit, than assuming a linear model with $\lambda=1$. Since other regressors should be taken into account however, a log-log regression between the relative innovation flows and relative cumulative flows is more practical. Table \ref{tab:logs2} reports the results. Previous innovation flows have a large predictive power of $56.5\%$ of the variation in the size of innovation flows.  
A one percent increase in previous ties increases the number of innovations with $0.675\%$. Meanwhile, underlying proximities have negative effects on the (log) size of network ties, explaining together $15.3\%$ of the variation between and within industries. This result lies in line with the notion that radical innovation activity is likely to be associated with unrelated variety, exploration and diversification. When fixed effects are used for sender and receiver nodes, such underlying proximities however explain little of temporal variation. Comparing results between Tables \ref{tab:logit}  and \ref{tab:logs} motivates the interpretation that underlying proximities are positively associated with the presence of innovation ties between industries, while smaller innovation flows are more common between related industries as measured through skill-relatedness and input-output flows. 

\subsection*{Similarity metrics and link prediction}
\label{sec:link}

\begin{figure*}
\caption{Boxplots of link prediction diagnostics}
\begin{subfigure}[b]{0.50\textwidth}
\caption{Accuracy of node similarity metrics, 1971-2013}
\label{fig:accuracy}
\centering
\includegraphics[width=8cm, height = 8cm, keepaspectratio]{accuracy.png}
\end{subfigure}
\begin{subfigure}[b]{0.50\textwidth}
\caption{Precision of node similarity metrics, 1971-2013}
\label{fig:precision}
\centering
\includegraphics[width=8cm, height = 8cm, keepaspectratio]{precision.png}
\end{subfigure}
\end{figure*}

\begin{table*}
\caption{Regressions with Katz metric}
\label{tab:katz}
\footnotesize
\centering
\begin{tabular}{lccccc} \hline
 & (1) & (2) & (3) & (4) & (5) \\
 &  &  &  &  & Fixed effects \\
VARIABLES & Base model & Logit & Log-log & Separate (log-log) & Separate \\ \hline
 &  &  &  &  &  \\
Katz $\beta = 20$ & 0.0275*** & 0.290***  & 0.722***  &  &  \\
 & (0.00143) &  (0.00914)&(0.00747)  &  &  \\
Katz, indirect ties &  &  &  & 0.0434*** & 0.0266*** \\
 &  &  &  & (0.00849) & (0.00967) \\
Direct ties &  &  &  & 0.648*** & 0.114*** \\
 &  &  &  & (0.00885) & (0.0138) \\
Constant & 3.40e-05*** & -0.697*** & -3.121*** & -1.070*** & -3.350*** \\
 & (4.60e-06) & (0.0453) & (0.0359) & (0.0593) & (0.264) \\
 &  &  &  &  &  \\
Observations & 273,910 & 59,024 & 5,816 & 5,816 & 5,816 \\
 R-squared & 0.145 & 0.0324 & 0.572 & 0.562 & 0.817 \\ \hline
\multicolumn{6}{c}{ Robust standard errors in parentheses} \\
\multicolumn{6}{c}{ *** p$<$0.01, ** p$<$0.05, * p$<$0.1} \\
\end{tabular}
\end{table*}

Alternative hypotheses to the preferential attachment mechanism departs from the notion of triadic closure and a role played by global network topology taking into account indirect paths between nodes. Table \ref{tab:corrtable} compares the relative share of innovations $\dfrac{\Delta w_{ij}}{\Delta w}$ with different similarity measures. It is clear that the best performing measures are the preferential weight assignment process and the Katz metric (coefficient $\beta=20$), taking into account indirect ties. Local metrics (Jaccard and Adamic-Adar) have low correlation. A complementary approach is to evaluate the predictors by using the $N$ highest ranked pairs of nodes according to each predictor, letting $N$ be equal to the number of innovations in a given calendar year. This analysis is evaluated in terms of measures of accuracy and precision. Accuracy is defined as the share of true positives (TP) and true negatives (TN) in the total number of possible edges: $\text{accuracy}=\dfrac{\text{TP}+\text{TN}}{\text{TP}+\text{TN}+\text{FP}+\text{FN}}$. Precision is defined as the ratio of true positives to all edges predicted by the similarity metric (false and true positives): $\text{precision}=\dfrac{\text{TP}}{\text{FP}+\text{TP}}$. Figures \ref{fig:accuracy} and \ref{fig:precision} summarize the performance of similarity metrics for the years 1971-2013. It is clear that preferential weight assignment and the Katz measure perform better than other potential link formation models. 

The Katz metric is further analyzed in Table \ref{tab:katz}. The Katz metric adds some information to the pure preferential attachment process in the logistic and log-log estimations (Tables \ref{tab:logit} and \ref{tab:logs}). When separating out indirect ties, equivalent to $S^{\text{katz}}_{ij}-\beta \dfrac{w_{ij}}{\sum w_{ij}}$, from direct ties $\dfrac{w_{ij}}{\sum w_{ij}}$, the elasticities of Model 4 suggest that indirect ties play a significant but smaller role. For a one percent change in indirect ties relative innovation flows increase with $0.04 \%$ while a one percent increase in direct ties produces an increase of $0.65 \%$.

\section{Conclusions}

The dynamic evolution of innovation systems is a process that arises from a non-trivial interaction of societal, economic and technological incentives, demands and opportunities. Understanding how technologies co-evolve and provide downstream and upstream stimulus for innovation elsewhere in the technological system requires quantitative and qualitative evidence. This study asked whether network stimulus matter in a quantitative sense for the significant innovations, and what factors determine the evolution of networks of significant innovations. 
 
This study finds evidence that significant innovations are afflicted by network stimulus, both from forward and backward linkages, in line with the received literature from innovation studies. Innovation network stimulus hence have a sizeable explanatory power of innovation patterns. Our results differ from those reported by \citet{acemoglu2016} since backward linkages have a pronounced role. This result is significant, since it is consistent with the view \citep{Rosenberg1969} that innovations upstream in technological systems target problems and imbalances downstream, also documented in earlier research on this database \citep{taalbi2017, taalbi2017origins}. One must note however that the network stimulus is based on a domestic flow, which is likely to underrate the role of supply of new opportunities.  

Previous research on this database has given qualitative evidence from innovation biographies of a dynamics where innovations are developed as a creative response to opportunities upstream or problems or imbalances downstream in the technological system in closely connected communities. The Swedish innovation network clearly suggests that the linkage formation of innovation networks is primarily shaped by historical network topology. The innovation network has a hierarchical structure with strong persistence and stability in network ties, forming through a preferential weight assignment process. Main supplier industries act as hubs, connecting diverse industries in the network. Our results hence speak in favor of the presence of key industries, some of which based on general-purpose technologies, creating a persistent and locally star-like community structure. In fact, most of these key industries are found in ICTs (compare \citealp{taalbi2017origins}). The evidence for exogenous factors was less compelling. While gravitational pull from important industries have consistent explanatory power, economic and skill relatedness displayed a weaker and more complex relationship to innovation flows.

These results have corollaries for our understanding of how technology shifts take place. First, our findings support the notions of network stimulus, preferential assignment and historical proximities as important mechanisms in the evolution of innovation systems and the formation of community structure. Our results, together with earlier research, should be taken to suggest that policy makers should monitor subsystems, centered on hubs and in driven by the opportunities upstream or imbalances downstream that focus innovation activity in the evolution of technological systems. 

Secondly, the results of this study suggest that, in terms of network link formation, networks of significant innovations appear to be a markedly different story from networks of regular invention activity. The mixed or negative association with conventional measures of economic, knowledge and skill proximities suggests that networks of radical innovations are driven by technological diversification, and only to a lesser extent correlated with inter-industrial proximities. Hence our study raises the question whether various proxies of innovation networks and technological interdependencies, e.g., constructed through imputed R\&D flows or patent citation networks, can be taken to reflect all types of innovation activity. While so, disruptive innovations that drive technological systems are not unpredictable as previous innovation ties between industries are strong predictors of future ties. Hence, understanding the evolution of technological systems hence requires the continued collection and analysis of direct measurements of innovation flows or patent network data that differentiate between radical and incremental innovation.  

\bibliographystyle{agsm}

\newpage
\clearpage
 \renewcommand{\theequation}{A.\arabic{equation}}
\setcounter{equation}{0}
\renewcommand*\thetable{A.\arabic{table}}
\setcounter{table}{0}
\renewcommand*\thefigure{A.\arabic{figure}}
\setcounter{figure}{0}

\appendix

\onecolumn
\section{Supplementary figures and tables}
\label{sec:network}

\begin{figure}[htpb]
\caption{\textbf{Innovation network 1970-2013. Relative flows between 29 sectors.}}
\label{fig:heatmap1}
\centering
\includegraphics[width=16cm, height = 18cm, keepaspectratio]{levelplot.png}
\end{figure}

\begin{figure*}
\caption{\textbf{Innovation network 1970-1989. Relative flows between 29 sectors.}}
\label{fig:heatmap2}
\centering
\includegraphics[width=16cm, height = 18cm, keepaspectratio]{levelplot7089.png}
\end{figure*}

\begin{figure*}
\caption{\textbf{Innovation network 1990-2013. Relative flows between 29 sectors.}}
\label{fig:heatmap3}
\centering
\includegraphics[width=16cm, height = 18cm, keepaspectratio]{levelplot9013.png}
\end{figure*}

\begin{table*}[htbp]
\caption{Correlation structure of network, five-year period and previous period}
\label{tab:correlation}
\centering
\footnotesize
\begin{tabular}{lc}
\hline
Period & Pearson corr.\\
\hline
1975-1979 & 0.62  \\& (0.00) \\
1980-1984 & 0.60  \\ &(0.00) \\
1985-1989 & 0.60 \\  & (0.00) \\
1990-1994 & 0.48 \\ & (0.00) \\
1995-1999 & 0.50 \\ & (0.00) \\
2000-2004 & 0.58 \\& (0.00) \\
2005-2009 & 0.62 \\ & (0.00) \\
2010-2013 & 0.46 \\& (0.00) \\
\hline
\end{tabular}
\end{table*}

\begin{table*} [htbp] \centering
\caption{Cross-correlation. Independent variables}
\label{tab:corrind}
\footnotesize
\begin{tabular}{l  c  c  c  c  c  c }\hline\hline
\multicolumn{1}{c}{Variables} &Cum. flows &Los &Skill-rel. &Gravity&Leontief &R\&D \\ \hline
Cumulative flows &1.0000\\
 &\\
Los index &-0.0282&1.0000\\
&(0.0000) &\\
Skill-rel.&0.0202&0.0711&1.0000\\
&(0.0000)&(0.0000) &\\
Gravity&0.0867&0.2565&-0.0030&1.0000\\
&(0.0000)&(0.0000)&(0.4230) &\\
Leontief &0.0176&0.0493&-0.0004&0.0589&1.0000\\
&(0.0000)&(0.0000)&(0.9108)&(0.0000) &\\
R\&D flows &0.0175&0.2217&0.1699&0.1444&0.0709&1.0000\\
&(0.0000)&(0.0000)&(0.0000)&(0.0000)&(0.0000)\\
\hline \hline 
 \end{tabular}
\end{table*}

\begin{table*}[htbp]\centering \caption{Cross-correlation table. Relative innovation flows and similarity metrics (lagged). \label{tab:corrtable}}
\footnotesize
\begin{tabular}{l  c  c  c  c  c  c }\hline\hline
\multicolumn{1}{c}{Variables} &Relative &Jaccard &Adamic-Adar &Katz & PWA & PA\\ 
\hline
Relative &1.0000\\
 &\\
Jaccard &0.0262&1.0000\\
&(0.0000) &\\
Adamic-Adar &0.0318&0.8044&1.0000\\
&(0.0000)&(0.0000) &\\
Katz $\beta=20$&0.3828&0.0556&0.0897&1.0000\\
&(0.0000)&(0.0000)&(0.0000) &\\
Pref. W.A. &0.3897&0.0527&0.0824&0.9801&1.0000\\
&(0.0000)&(0.0000)&(0.0000)&(0.0000) &\\
Pref. Att. &0.2133&0.0833&0.1628&0.5340&0.4650&1.0000\\
&(0.0000)&(0.0000)&(0.0000)&(0.0000)&(0.0000)\\
\hline \hline 
 \end{tabular}
\end{table*}

 \renewcommand{\theequation}{B.\arabic{equation}}
\setcounter{equation}{0}
\renewcommand*\thetable{B.\arabic{table}}
\setcounter{table}{0}
\renewcommand*\thefigure{B.\arabic{figure}}
\setcounter{figure}{0}
\clearpage
\twocolumn
\section{Weight distribution}
\label{sec:distribution}
The network studied in this paper can be formalized as a network where one innovation is added at each point in time $t$, such that the sum total of innovations is $t$.
It is also the case that the number of edges actually used is only a fraction of the total possible. With this in mind, it is possible to specify a model in which time $t$ is smaller than the (fixed) number of edges $A$ out of which $N$ have non-zero weight. In general, following \citet{krapivsky2001}, with an attachment kernel $a_w$ that governs the flow of innovations between industries, a rate of weight assignment to unused edges $\beta$ and an appropriate normalization factor $M = \sum_w a_w N_w$, with $N_w$ being the number of edges with $w$ innovations, the master equation is

\begin{equation}
\dfrac{\partial N_w}{\partial t} = \beta P(w) = \dfrac{a_{w-1}}{M} N_{w-1} - \dfrac{a_{w}}{M} N_{w}
\end{equation} which can be rearranged into 

\begin{equation}
P(w)= \dfrac{a_{w-1}}{\mu \beta + a_w} P(w-1)
\end{equation} where $\mu = M/N$ and $P(w) = N_w /N$. 
 Substituting $P(w-1)$ recursively, it follows that

\begin{equation}
\label{eq:master}
P(w) =\dfrac{1}{a_w} \prod_{j=1}^w \left(\dfrac{\beta \mu } {a_j}+1\right)^{-1}
\end{equation}

A model of preferential attachment with heterogeneity specifies that the probability of an innovation flowing between industry $i$ and $j$ is related to weight of the edge $w_{ij}/\sum w_{ij}$. In practice, it is possible to model the evolution of the innovation network as a mixed process or random and preferential attachment to edges. With probability $\alpha$, innovations flow between industry pairs $i,j$ proportional to $\dfrac{w^\lambda}{\sum_w w^\lambda N_w}$. Innovations also appear randomly in new industry pairs $i,j$ with probability $1-\alpha$. Hence, the probability of an innovation to be added to an (ordered) industry pair with previous weight $w$ is written as

\begin{equation}
\Pi_w = (1-\alpha) \delta_{w,0} +\alpha \dfrac{w^\lambda}{\sum_w w^\lambda N_w}
\end{equation}

For the empirical case where the number of edges is fixed, this equation can be motivated as an average over a process where edges with weight zero have probability $(1-\alpha) = N/A$. 

Using the master equation, and $P(w) = N_w/N$ and $\partial N/\partial t = m(1-\alpha)$, where $m$ is the number of links (user industries) each innovation has,

\begin{equation}
P(w) = \dfrac{1}{w^\lambda} \prod_{j=1}^w \dfrac{j^\lambda}{\dfrac{(1-\alpha) m \mu}{\alpha} +j^\lambda}
\end{equation} For $\lambda=1$, one may use the Gamma function $\Gamma(x+1)=x!$ to derive

\begin{multline} \label{eq:linear} P(w) \propto w^{-1}\dfrac{\Gamma (w+1)}{\Gamma (\dfrac{(1-\alpha) m \mu }{\alpha}+w+1)} \\ \sim w^{-1-1/\alpha} 
\end{multline} since for $\lambda=1$, $\mu = \sum_w w P(w) = \dfrac{t}{(1-\alpha) m t}= \dfrac{1}{(1-\alpha) m }$. 

For $\lambda<1$, the product in equation \ref{eq:master} is written as the exponential of a sum. Using the series representation of $\ln \left(\dfrac{(1-\alpha) m \mu(\lambda)}{\alpha}w^{-\lambda} +1 \right)$ and integrating over $w$ yields

\begin{multline}
P(w) \propto \\ w^{-\lambda} \exp - \sum_{n=1}^{\infty} \dfrac{(-1)^n}{n}  \left( \dfrac{(1-\alpha) m \mu(\lambda)}{\alpha} \right)^n \dfrac{w^{1-\lambda n}}{1-\lambda n}
\end{multline} which for $\dfrac{1}{2}\leq \lambda <1$ is 

\begin{equation}
\label{eq:sublinear}
P(w) \propto w^{-\lambda} \exp - \left( \dfrac{(1-\alpha) m \mu(\lambda)}{\alpha} \right) \dfrac{w^{1-\lambda}}{1-\lambda } 
\end{equation} The cumulative weight distributions in the linear and sub-linear cases (Figure \ref{fig:weights}) are obtained by integrating equations \ref{eq:linear} and \ref{eq:sublinear}

\begin{equation}
\int w^{-1-1/\alpha} dw = w^{-1/\alpha}
\end{equation}

and

\begin{multline}
\int w^{-\lambda} \exp - \left( \dfrac{(1-\alpha) m \mu(\lambda)}{\alpha} \right) \dfrac{w^{1-\lambda}}{1-\lambda }  dw \\
\propto  \exp - \left( \dfrac{(1-\alpha) m \mu(\lambda)}{\alpha} \right) \dfrac{w^{1-\lambda}}{1-\lambda }
\end{multline} The parameters of the linear model are given by estimating Model 1 in Table \ref{tab:results}. The non-linear, in practice sub-linear, model is estimated in two steps in Table \ref{tab:logs}. $\lambda$ is estimated to 0.65. Using the fact that every innovation on average has $m=2.44$ user industries, $\dfrac{(1-\alpha)}{\alpha} \mu$ is estimated to 0.596 in simulations. The resulting distribution is shown in Figure \ref{fig:weights}.

\end{document}